\documentstyle{mn}

\begin{document}
                                                                                
\title[A complete sample of radio sources -- III]
{A complete sample of radio sources in the North Ecliptic
Cap, selected at 38 MHz -- III. further imaging observations and 
the photometric properties of the sample}

\author[Lacy et al.]{Mark Lacy$^{1}$\thanks{Guest observer, McDonald
Observatory, University of Texas at Austin}, Mary Elizabeth Kaiser$^{2,3}$, 
Gary J. Hill$^2$, Steve Rawlings$^{1}$ \\
{\LARGE \& Gareth Leyshon$^{4}$}\\
\\
$^{1}$Astrophysics, Department of Physics, Keble Road, Oxford, OX1 3RH.\\
$^{2}$McDonald Observatory, University of Texas at Austin, TX78712-1083\\
$^{3}$Bloomberg Center for Physics and Astronomy, Johns Hopkins University, 
3400 Charles Street, Baltimore, MD21218 \\
$^{4}$Astronomy Dept, University of Wales College of Cardiff PO Box 913, 
Cardiff CF2 3YB}

\maketitle
\begin{abstract}
Further imaging observations of a sample of radio sources in the North
Ecliptic Cap are presented and a number of new identifications are made. 
Using redshifts from spectroscopic data 
presented in a companion paper (Lacy et al.\ 1999b), the photometric
properties of the galaxies in the sample are discussed. It is shown
that: (1) out to at least $z\approx 0.6$ radio galaxies are good
standard candles irrespective of radio luminosity; (2) for 
$0.6\stackrel{<}{_{\sim}} z \stackrel{<}{_{\sim}} 1$ a large fraction
of the sample have magnitudes and colours consistent with a
non-evolving giant elliptical, and (3) at higher redshifts, where the 
$R$-band samples the rest-frame UV flux, most objects have 
less UV luminosity than expected if they form their stellar populations 
at a constant rate from a high redshift to $z\sim 1$ in unobscured star-forming
regions (assuming an Einstein -- de Sitter cosmology). The
consequences of these observations are briefly discussed.

\end{abstract}


\section{Introduction}

Radio galaxies are at present the only high redshift galaxies which
are not explicitly selected on the basis of high star formation rates, 
but which can nevertheless be found in reasonably large numbers. 
This makes
them potentially very valuable for understanding the evolution of 
massive galaxies at early times. Unfortunately, the strong correlation
of the radio and UV emission in the most radio-luminous
galaxies (the ``alignment effect''; see, e.g., Best et al.\ 1997 and 
Ridgway \& Stockton 1997) indicates that 
their photometric 
properties are probably dominated by the presence of an AGN, at least at
rest wavelengths $<4000$\AA, a deduction which was confirmed by 
spectroscopically for most $z\sim 1$ 3C radio galaxies 
by Hammer, LeF\`{e}vre \& Angonin (1993). 
This makes defining and studying samples of 
radio sources selected at fainter flux density levels important,
as, due to the correlation of radio luminosity with the optical/UV
luminosity of the AGN (e.g.\ Willott et al.\ 1999), such objects should 
have only a very small fraction of their light produced by the AGN. 

In two previous papers (Lacy, Rawlings \& Warner 1992; 
Lacy et al.\ 1993; hereafter Papers I and II), we defined and presented 
radio and optical imaging data for a sample of radio sources selected
from the 38-MHz survey of Rees (1990) in the region of the North Ecliptic 
Cap (NEC).
In a companion paper (Lacy et al.\ 1999b; hereafter Paper IV) we 
present optical 
spectroscopy of this sample, along with that of a sample selected at 151MHz
with a comparable flux limit and in the same region of sky (the 7C-{\sc iii}
sample, Lacy et al.\ 1999c) and which therefore has most 
objects in common with this sample. 

We have already discussed the outcome of high-resolution optical 
observations of a $0.5<z<0.82$ sub-sample of objects
from the 7C-{\sc iii} sample which show that the radio -- 
optical ``alignment effect'' in the 7C-{\sc iii} sample is weak, but probably
present, even at wavelengths longward of 4000\AA$\;$(Lacy et al.\ 1999a). 
Thus is it possible that even in these radio-faint ($\approx 20$-times fainter
than 3C) samples, the radio source is influencing the properties of 
the host galaxy. As we discuss
there, however, the scale of the alignments does seem to change with radio
luminosity, in the sense that the alignments of the 7C-{\sc iii} radio 
galaxies are detectable only on
small scales (in $\sim 15$ kpc radius apertures), whereas the 3C radio 
galaxies have equally strong 
alignments at 15 kpc and 50 kpc. This could be because the highly 
radio-luminosity dependent 
mechanisms such as scattering of the hidden quasar nucleus and nebular 
emission disappear in the 7C-{\sc iii}
objects, leaving a less luminosity-dependent mechanism which only operates 
on small scales. Relatively ``passive''
mechanisms may be responsible for these alignments, for example dust 
disks with axes
along the radio axis and/or selection effects. If this were the case 
the photometric properties of the underlying galaxy would be unaffected by
the presence of the radio source. However, it is possible that jet-induced star
formation is only weakly radio-luminosity dependent, and if this mechanism
operates then it would imply an intimate connection 
between the presence of the radio source and the stellar population of the 
host. The observational evidence for this mechanism is as yet only 
circumstantial however, and it is not clear 
whether such a mechanism could be effective (Icke 1999).

In this paper we discuss further optical, radio and near-infrared
observations of the 8C sample, and use the spectroscopy presented in
Paper IV to discuss the photometric properties and evolution of the
radio source hosts. Unless otherwise stated we assume an $\Omega_{\rm M}=1,
\Omega_{\Lambda}=0$ (Einstein -- de Sitter) cosmology with 
$H_0=50 {\rm kms^{-1}Mpc^{-1}}$.

\section{observations}

Radio snapshot observations of two of the sources in the sample,
8C1804+632 and 8C1826+660 were made 
with the VLA in C-array at 1.4 GHz 
on 1993 June 28 to attempt to resolve ambiguities of 
identification in Paper II. 8C1742+637 was observed in A-array
at 1.4 GHz on 1995 June 29 for $2 \times 2$min snapshots. 
These data were calibrated
and analysed in the standard manner and the maps are presented in Fig.\ 1.
In addition a four minute 8GHz A-array observation of 8C 1826+651 was made on
1996 December 31 to 
attempt to identify a central component. A 0.6 mJy point source was detected 
at 18 26 31.41 +65 10 46.0 (B1950), this is consistent with the identification
in Paper II and corresponds to optical/infrared component `b' 
of Fig.\ 2 being the probable radio source host (see also Lacy et al.\ 1998a).

Observations of several sources were made in the near-infared 
$H$-band with ROKAM, 
a near-infrared camera with a $256 \times 256$ HgCdTe array, during 1993 May 
and June. ROCAM was placed at the 
f/18 straight Cass focus of the McDonald 2.7-m telescope, giving a pixel 
scale of 0.40 arcsec pixel$^{-1}$. Objects with estimated redshifts $\approx 1$
in Paper II were the primary targets, though some lower redshift objects 
were also observed when the seeing was very poor. The observations are 
detailed in Table 1,
and the photometry in Table 4. Greyscales of selected images are presented in 
Fig.\ 2. The observations were calibrated using the infrared standard stars 
HD136754 and HD162208 (Elias et al.\ 1982). Extinction was determined 
to be $0.15 \times {\rm airmass}$ and a correction of 0.05 
magnitudes was made for galactic extinction,
assuming $E(B-V) = 0.085$ (Kolman et al.\ 1991) and the extinction curve of 
Cardelli, Clayton \& Mathis (1992). 

Further $R$ and $I$-band observations of the unidentified sources in the 
sample were made on the McDonald 2.7-m using IGI (see Paper II) 
and on the William Herschel Telescope on La Palma using the Auxilliary Port.
Details of these observations are given in Table 2. The extinction assumed 
at the WHT in July was $0.15 \times {\rm airmass}$ in $R$, with a galactic
extinction of 0.19 and the data were 
calibrated in the standard star PG 2331+055. In $I$-band, an 
extinction of $0.02 \times {\rm airmass}$ was assumed for the 
observations made in 1994 January, and $0.05 \times {\rm airmass}$ in 
1994 July (higher due to Saharan dust). A value of $0.07 \times {\rm airmass}$
was assumed for the McDonald observations. Galactic extinction in $I$-band 
was estimated to be 0.13 mag using the extinction curve of Cardelli et al.\ 
(1992).

\begin{table}
\caption{$H$-band observations}
\begin{tabular}{ccclc}
Source & Date & Exposure time & airmass&Seeing \\
       &      &         /s    &        &/arcsec\\
8C1743+645 & 07/05/93 & 2970 &1.22 & 2.8 \\
           & 12/06/93 & 2970 &1.22 & 1.9 \\
           &          &      &     &     \\
8C1745+642 & 06/05/93 & 1890 &1.20 & 1.6 \\
           & 13/06/93 & 1620 &1.20 & 1.5 \\
           &          &      &     &     \\
8C1748+675 & 07/05/93 & 2160 &1.32 & 2.4 \\
           &          &      &     &      \\
8C1755+632 & 07/05/93 & 1890 & 2.0 & 2.7 \\
           &          &      &     &     \\
8C1755+685B& 07/05/93 & 1620 & 1.6 & 1.8 \\
           &          &      &     &     \\
8C1757+653 & 09/05/93 & 1890 & 1.32& 3.0 \\
           &          &      &     &     \\
8C1803+661 & 08/05/93 & 1800 & 1.28& 2.0 \\
           & 12/06/93 & 5490 & 1.24& 1.6 \\
           &          &      &     &     \\
8C1805+635 & 11/06/93 & 3780 & 1.25& 1.6 \\
           &          &      &     &     \\
8C1807+685 & 07/05/93 & 2700 & 1.29& 2.6 \\
           &          &      &     &     \\
8C1811+633 & 08/05/93 & 2520 & 2.0 & 2.3 \\
           &          &      &     &     \\
8C1815+682 & 09/05/93 & 1350 & 1.27& 2.0 \\
           & 12/06/93 & 3240 & 1.4 & 1.7 \\
           & 13/06/93 &  810 & 1.36& 1.5 \\
           &          &      &     &      \\
8C1816+671 & 09/05/93 & 2610 & 1.28& 2.3 \\
           &          &      &     &      \\
8C1821+643 & 08/05/93 & 1080 & 1.4 & 3.2  \\
           &          &      &     &      \\
8C1823+660 & 08/05/93 & 1980 & 1.6 & 3.5  \\
           &          &      &     &      \\
8C1826+651 & 06/05/93 & 1620 & 1.22& 1.6  \\
           &          &      &     &      \\
8C1827+671 & 11/06/93 & 1080 & 1.4 & 1.5  \\
	   & 13/06/93 & 5670 & 1.25& 1.4  \\
\end{tabular}
\end{table}

\begin{table*}
\caption{Further optical observations}
\begin{tabular}{llccrlc}
Source &Telescope/instrument& Filter & Date &Exposure time &airmass& seeing \\
       &                    &        &      &    /s         &      & /arcsec \\
8C 1743+645 & McDonald 2.7-m/IGI & $R$ & 21/5/93 & 1200 & 1.20 & 2.0 \\
8C 1748+670 & WHT/Aux Port (TEK) & $R$ & 31/7/95 &  600 & 1.8  & 0.8   \\ 
8C 1753+664 & WHT/Aux Port (TEK) & $I$ & 25/1/95 &  900 & 1.7  & 0.8 \\
8C 1754+643 & WHT/Aux Port (TEK) & $I$ & 26/1/95 &  900 & 1.6  & 2.0 \\
8C 1757+653 & McDonald 2.7-m/IGI & $I$ & 21/5/93 & 3300 & 1.7  & 1.9  \\
8C 1757+653 & WHT/Aux Port (TEK) & $I$ & 25/1/95 &  600 & 1.6  & 0.8  \\
8C 1814+670 & WHT/Aux Port (TEK) & $I$ & 25/1/95 &  900 & 1.7  & 1.0 \\
8C 1816+671 & McDonald 2.7-m/IGI & $I$ & 21/5/93 & 2400 & 1.6  & 1.5  \\
8C 1826+660A& WHT/Aux Port (TEK) & $R$ & 30/7/95 &  900 & 1.8  & 0.8 \\
\end{tabular}
\end{table*}

\begin{table*} 
\caption{Additions to and new identifications for objects in the 8C sample}
\begin{tabular}{lrrrll}
Name & R.A.\ (1950) & Dec.\ (1950) & $R$-magnitude &Type & Notes\\
8C1732+672 &   -    &     -        &             &?& Addition, but badly 
confused in radio so excluded from complete sample\\
8C1733+673 & 17 32 59.81&+67 19 32.3& 22.6      &G&Addition, $R$-magnitude 
from spectrum\\
8C1736+650 & 17 36 26.61&+65 04 10.7& 23.0      &G&Misidentification in Paper
II, position uncertain, $R$ from spectrum\\
8C1742+637 & 17 42 43.39&+63 46 11.8& 22.6       &G&Addition, identification assmued to be `a' in Fig.\ 3\\
8C1743+645 & 17 43 28.55&+64 31 30.7& 23.3       &G&Misidentification in 
Paper II  \\
8C1748+670 & 17 48 12.90&+67 03 54.2& 23.9       &G&Misidentification in 
Paper II ?\\
8C1754+643 & 17 54 11.89 &+64 20 36.5& $>$23.1   &G&Misidentification in 
Paper II. $I=$21.1\\   
8C1804+632 & 18 04 49.37 & +63 13 07.9&$\sim 23$&G&New identification; see Section 4\\
8C1805+635 & 18 05 37.33 &+63 32 47.8& $>$23.7   &G&Misidentification in Paper
II\\
8C1814+670 & 18 14 16.09 &+67 02 44.5& $>$22.9  &G&New identification 
($I=23.4$) \\
8C1816+671 & 18 16 30.40 &+67 11 00.2& $>$23.8   &G&New identification 
($I=22.4$)\\
8C1826+660A& 18 25 47.54 &+66 02 24.0& 23.8      &G&New identification; note 
the finding chart in Paper II is incorrect\\
8C1827+652 & 18 27 59.13&+65 17 49.4&  -        &G?&Addition, but near bright
stars so excluded from complete sample\\
\end{tabular}

\vspace*{0.1in}

\noindent
Notes: This Table supplements table 2 of Paper II. 
Positions are the positions of the optical identifications (except for 
8C1827+652 where it is the radio central component) and 
should be accurate to $\stackrel{<}{_{\sim}} 1.5$ arcsec, except for 
8C1736+650 whose identification is roughly at the mid-point of the radio 
hotspots, but is only seen in our spectrum and so the uncertainty is larger. 
Magnitudes are measured in the standard metric aperture of 63.9 kpc, if no 
redshift is known $z=1$ is assumed.
\end{table*}

\begin{figure}
\setlength{\unitlength}{1mm}
\begin{picture}(80,170)
\put(-20,-90){\includegraphics{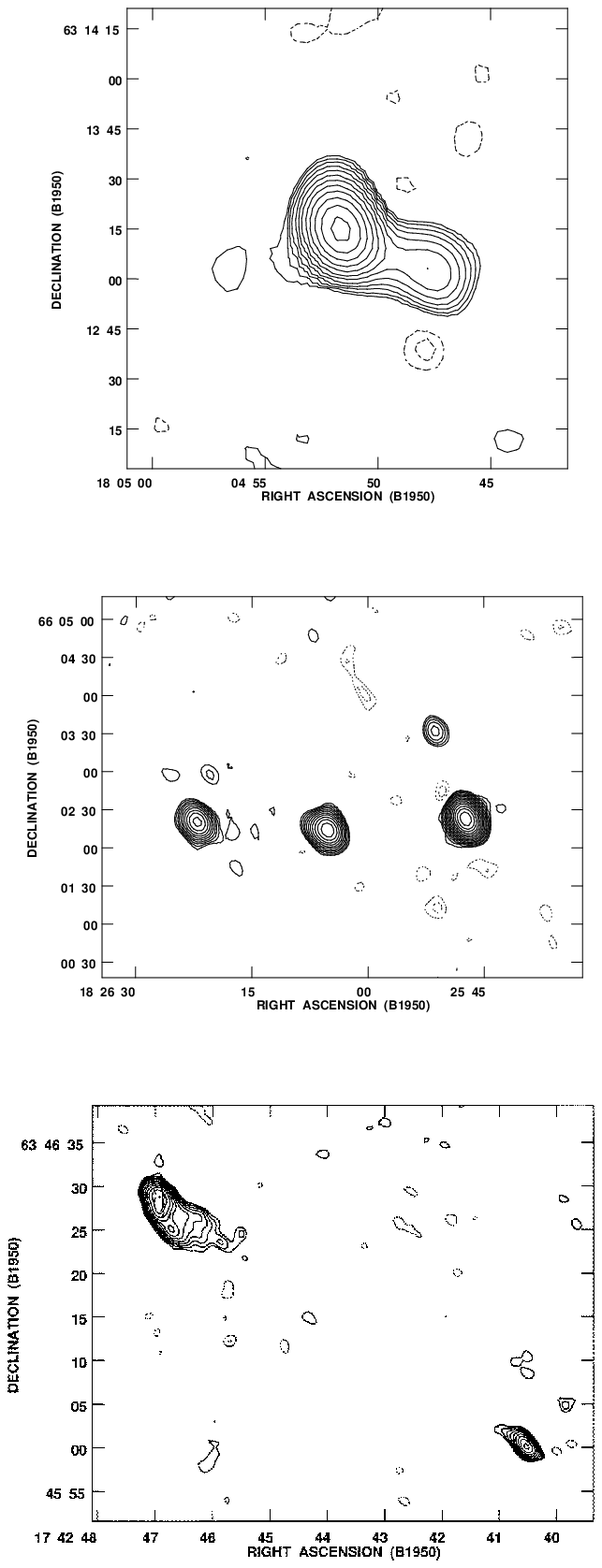}}
\end{picture}
\caption{New radio observations of sources in the 8C sample at 1.4 GHz:
(a) 8C1804+632, (b) 8C1826+660 and (c) 8C1742+637. Contours are 
logarithmic, spaced by
factors of two from $\pm 1$mJy/beam in (a) and (b), and by factors of 
$\sqrt{2}$ from 0.3 mJy/beam in (c). The restoring beam in (a) and (b) has 
FWHM $17^{''} \times 13^{''}$ at PA 35 deg., that in (c) has FWHM 
$1.8^{''} \times 1.2^{''}$ at PA 50 deg. The polarisation vectors in (c) are 
$E$-field and scaled such that 1-arcsec is equivalent to 1.1 mJy/beam. 
Negative contours are shown dashed.}
\end{figure}

\begin{figure*}
\setlength{\unitlength}{1mm}
\begin{picture}(160,130)
\put(-20,-80){\includegraphics{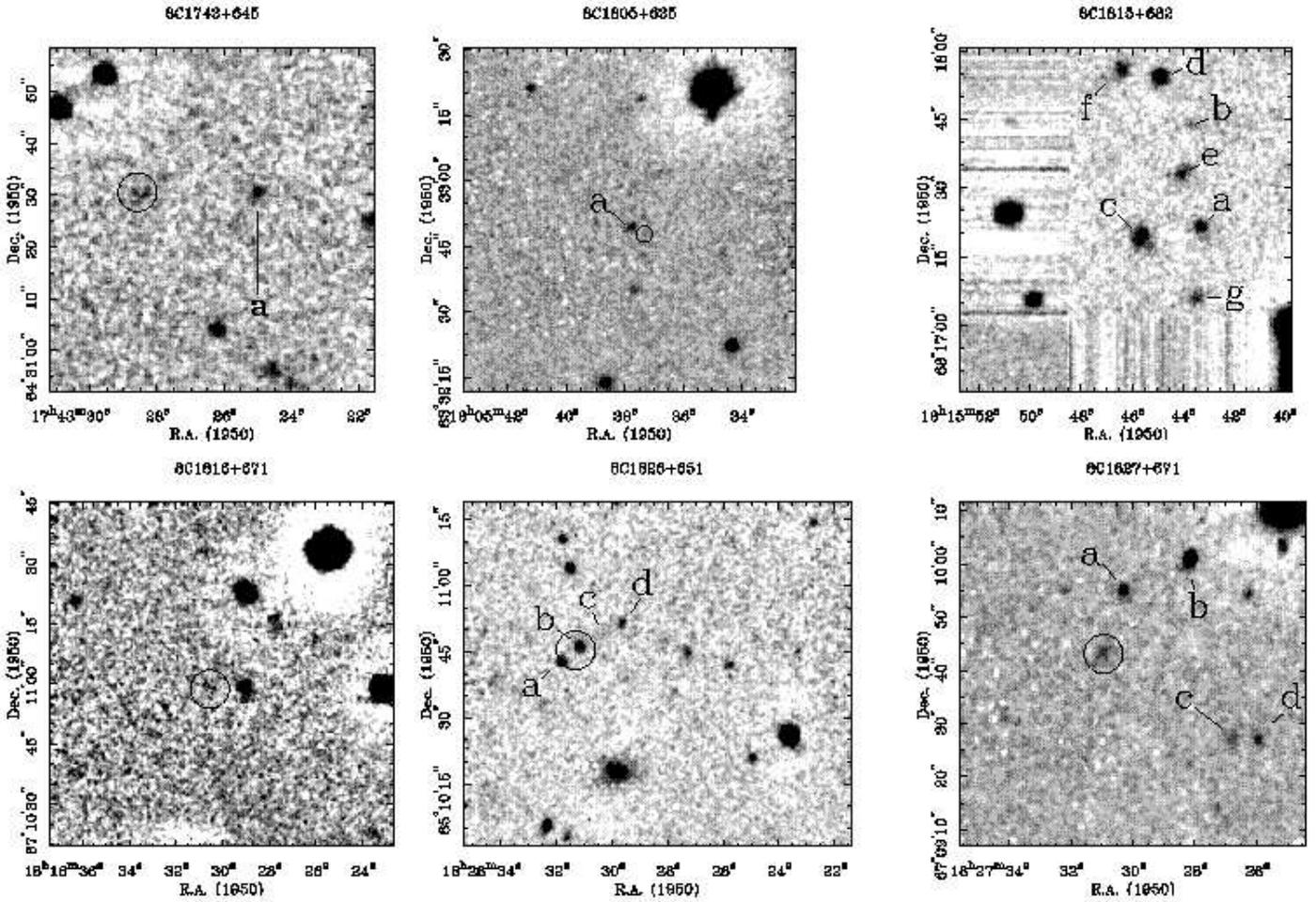}}
\end{picture}

\caption{Selected $H$-band observations of the sources in the 8C sample. The 
host galaxies are indicated by either a ring around the radio position of
Paper 1, or, where there are several candidate identifications, by lettering.
Candidate companion objects are also indicated by lettering in some cases.}
\end{figure*}




\begin{figure*}
\setlength{\unitlength}{1mm}
\begin{picture}(150,190)
\put(-30,-40){\includegraphics{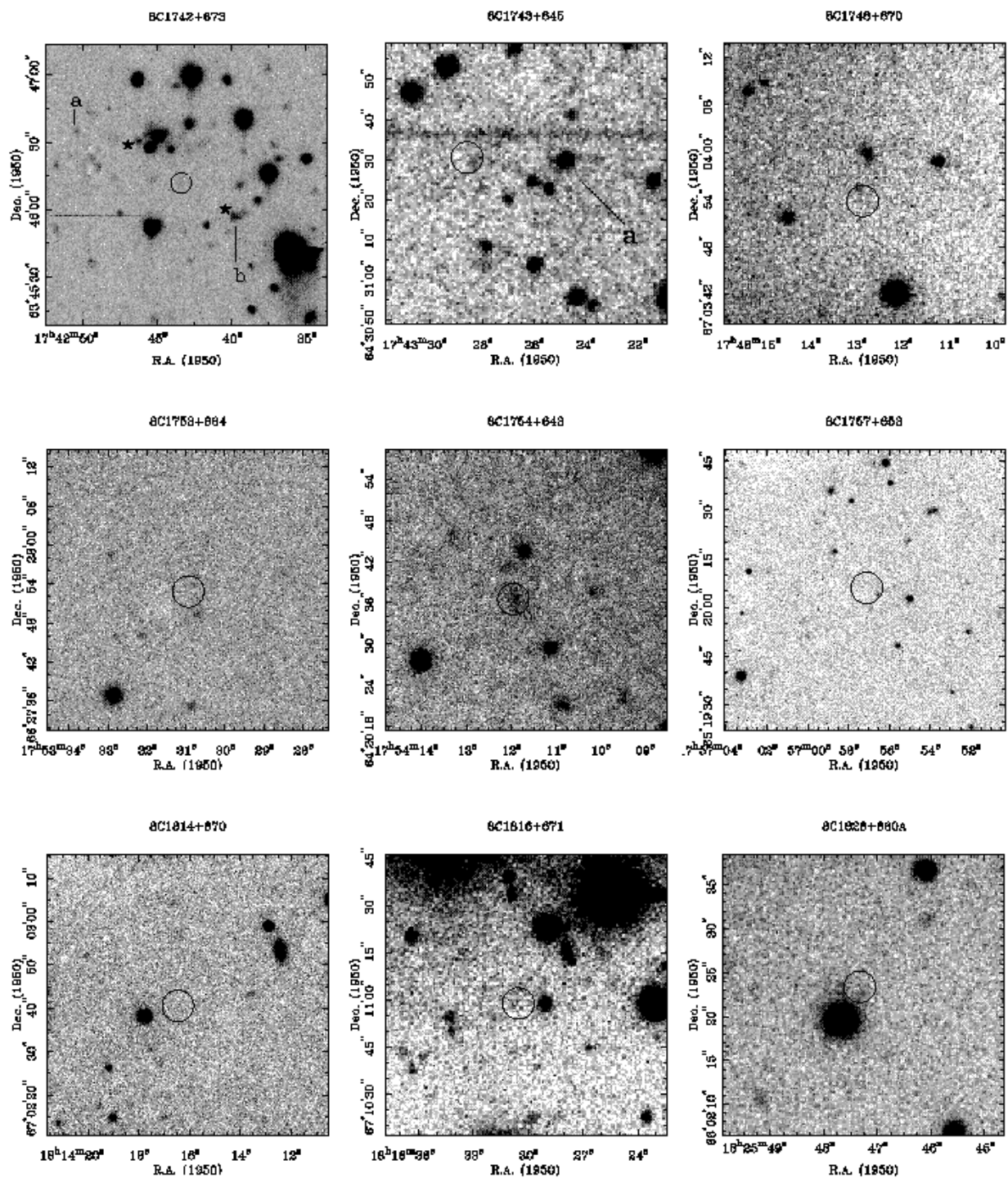}}
\end{picture}

\caption{New optical ($R$- or $I$-band) observations of sources in the 8C 
sample. The identification, or the best estimate of the radio source 
position where no identification exists, is indicated by the circle. 
The positions of the radio hotspots of 8C1742+637 are marked with stars.}
\end{figure*}

\begin{table}
\caption{$H$-band photometry of sources in the 8C sample}
\begin{tabular}{cccc}
Source & $H$-magnitude & aperture & $R-H$ \\
       &               & diameter &       \\
       &               & /arcsec  &       \\
       &               &          &       \\     
8C1743+645 & 18.8      &  5       & 4.7   \\
           &           &          &      \\
8C1745+642 & 17.5      &  6       & 3.2  \\
object `a'&  18.9     &   5      &  $>4.2$\\
object `b'&  17.9     &   5      & 3.4 \\
           &     &     &       \\
8C1748+675a&16.4 & 8.2 &3.5 \\
8C1748+675b&17.8 & 6   &3.1\\
8C1748+675c&17.9 & 6   &3.6\\ 
           &     &     &    \\
8C1755+632 & 15.3&18 & 2.6\\
           &    &     &     \\
8C1755+685B&16.9& 6 & 3.2 \\
           &    &     &     \\
8C1757+653 &$>19.4$ &5 &-\\
           &      &     &     \\
8C1803+661 &18.3 &7.1 &2.8 \\
          &      &     &     \\
8C1805+635 &$>20.7$ & 5 & - \\
object `a' &18.5 & 4.6 & 2.6\\
           &      &     &     \\
8C1807+685 &16.5 & 6& 3.7 \\
           &      &     &     \\
8C1811+633 & 16.2& 7.1 & 3.1 \\
           &      &     &     \\
8C1815+682a& 17.8 & 6 & 3.3\\
8C1815+682b& 19.6 & 4.6 & 1.7 \\
8C1815+682c& 17.1 & 6  & 3.8 \\
8C1815+682d& 17.0 & 6  & 3.0 \\
8C1815+682e& 17.9 & 6  & 3.8 \\
8C1815+682f& 17.7 & 6  & 3.7 \\
8C1815+682g& 18.3 & 4.6& 3.4 \\
           &      &     &      \\
8C1816+671 &19.0 & 5& $>$5.0 \\
           &      &     &      \\
8C1821+643 & 12.6 & 8 & -  \\
           &      &     &      \\
8C1823+660 & 15.8 & 12 & 2.9 \\
           &      &     &      \\
8C1826+651a&17.0 & 5 & 2.2 \\
8C1826+651b&17.1 & 5 & 3.8 \\
8C1826+651c&18.4 & 5 & 3.4 \\
8C1826+651d&17.9 & 5 & 3.5 \\
           &      &     &      \\
8C1827+671 & 18.5 & 8.2 & 2.6  \\
object `a' & 18.3 & 5  & 3.8  \\
object `b' & 17.1 & 8.2 & 3.1 \\
object `c' & 19.2 & 5 & $>$4.3 \\
object `d' & 18.9 & 5 & $>$4.6 \\
           &      &   &       \\
\end{tabular}

\end{table}

\begin{table}
\caption{Further optical photometry of sources in the 
8C sample}
\begin{tabular}{llcc}
Source & magnitude & aperture & $R-I$ \\
       &               & diameter &       \\
       &               & /arcsec  &       \\
       &               &          &        \\
8C1743+645 & $R=23.5$  & 5     & - \\
8C1748+670 & $R=24.2$  & 3     & - \\
8C1753+664 & $I=23.2$ ($3.0 \sigma$) & 5 & $>0.4$\\
8C1754+643 & $I=21.3$  & 5 & $>1.3$ \\
8C1757+653 & $I>23.1$  &5& - \\
8C1814+670 & $I=24.0$ ($3.0 \sigma$) &2& $>-0.8$ \\
8C1816+671 & $I=22.6$ ($3.7 \sigma$)&5 & $>2.6$\\ 
8C1826+660A& $R=24.0$ &2& - \\
\end{tabular}
\end{table}




\section{Additions to the sample}

Careful study of a 7C survey of the NEC (Lacy et al.\ 1995a) revealed 
four objects which, although absent from the original sample of 
Paper I due to 
confusion, should probably be included in the complete sample. 
Two of these remain excluded due to radio or optical confusion as 
discussed below. A full 
description of the objects including further radio maps and finding 
charts will be presented in Lacy et al.\ 1999c, 
but is summarised here and in Table 3 for completeness: 

\subsection*{8C 1732+672 (7C 1742+6715)}

The 1.5 GHz VLA map of this source is very confused, and so it has 
been temporarily removed from the sample again pending improved radio data.

\subsection*{8C 1733+673 (7C 1733+6719)}

This is a small ($2.5^{''}$) apparently double radio source, PA $15^{\circ}$.
It is identified with 
an $R\approx 22.6$ galaxy (measured from the optical spectrum and corrected 
to the standard metric aperture).

\subsection*{8C 1742+637 (7C 1742+6346)}

An FRII radio source with a marginal detection of a possible central 
component at 17 42 44.36 +63 46 15.0 (B1950) which is  
$51^{''}$ in size (Fig.\ 1). The position of the nearest likely 
optical identification, ringed in Fig.\ 3, is given in Table 3. It lies 
along the radio axis, but not coincident with the putative central component, 
and has an $R$-magnitude of 22.6 in the 
standard metric aperture. Both the hotspots are projected close to 
what are probably foreground galaxies (the source lies 4.5 arcmin 
from the cluster Abell 2280) so significant gravitational lensing may be 
occuring.

\subsection*{8C 1827+652 (7C 1827+6517)}

Another FRII radio source $17^{''}$ in size at PA $155^{\circ}$. This object 
is very close to several bright stars so has again been temporarily removed 
from the sample.

\section{Comments on individual objects}

\subsection*{8C1743+645}

The identification given in Paper II (`a' in Fig.\ 2) closely
coincides in position with the
eastern radio hotspot. The $H$-band image (Fig.\ 2) reveals a faint red
object coincident with the putative radio central component (Paper I)
which is a much more likely identification. This galaxy is also just 
visible in $R$-band (Fig.\ 3). Lensing of the East
hotspot by the (presumably) foreground galaxy `a' is likely to be occuring.

\subsection*{8C1748+670}

There are two candidate identifications, the first one was listed in Paper II, 
but our WHT $R$-band image (Fig.\ 3) also revealed a fainter object 1.4 
arcsec closer to the radio central component whose position is given in 
Table 3. If real, this is the more likely identification. The revised 
position of the object discussed in Paper II is 17 48 12.95 +67 03 55.6 
(B1950), the old position was in error, presumably due to the low signal:noise
of the detection.

\subsection*{8C1754+643}

The identification given in Paper II is incorrect; the true identification 
is a faint, diffuse galaxy located between the radio hotspots, ringed in 
Fig.\ 3.

\subsection*{8C1804+632}

As discussed further in Paper IV, 
despite the additional radio imaging presented in Fig.\ 1, the nature of
this radio source and its identification continues to elude us. The
fairly compact source detected at 5GHz in Paper I now appears as just
a part, perhaps a hotspot, of a larger, $\approx 30$\arcsec\ radio
source in our 20cm map. Spectra taken through objects `a', `F' and `b' 
of Fig.\ 4 showed that all three were probably stars. The most likely 
identification at present is object `c', which is unfortunately very close to 
star `F'. Consequently the magnitude of this object is very uncertain, but
we estimate $R \approx 23.6$ in a 2$^{''}$ diameter aperture.

\begin{figure}
\begin{picture}(200,200)
\put(0,230){\includegraphics{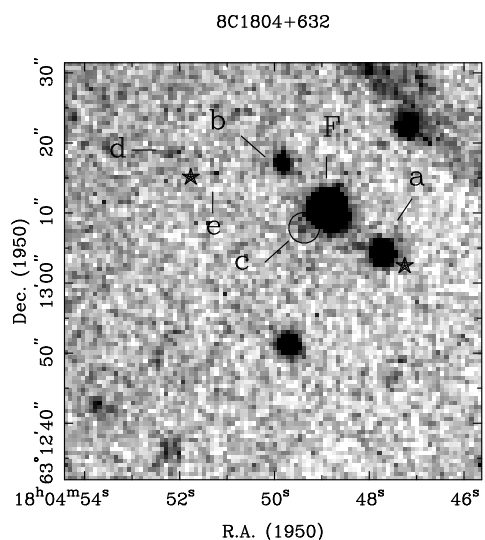}}
\end{picture}
\caption{Finding chart for 8C1804+632. The objects mentioned in the 
text are labelled and the positions of the presumed radio hotspots from Fig.\ 1
marked with stars.}
\end{figure}

\subsection*{8C1805+635}

The original identification of this object in Paper II
was an $R=21.1$ galaxy 3 arcsec from line joining the midpoint 
of the radio hotspots (`a' in Fig.\ 2). The spectroscopy of Paper IV 
shows that the true identification is on the axis, and that `a' is
foreground, and possibly lensing the radio emission.

\subsection*{8C1814+670}

There is a $3\sigma$ detection of a possible identification, 4.5 arcsec
NW from the mid-point of the radio hotspots (Fig.\ 3).

\subsection*{8C1816+671}

Another new identification, faintly visible in both $H$ (Fig.\ 2) and 
$R$ (Fig.\ 3).

\subsection*{8C1821+643}

Better known as the quasar E1821+643, this object is one of the most 
luminous radio-quiet quasars (RQQ) known with $z<0.5$, and thus has one of
the brightest radio fluxes of an RQQ. 
The optical, X-ray and 
near-infrared properties of the quasar have been 
studied by Kolman et al.\ (1993, 1991). The long-term variability study of 
Kolman et al.\ (1993) shows that the optical and UV flux from the quasar
declined between 1987 and 1991; our $H$-band data is consistent with a 
decline from 1988 April 25, when Kolman et al.\ measured $H=12.19$, to 
our observations in 1993 May 8 ($H=12.6$), although due to the errors in the
photometry this result should be considered tentative.

\subsection{8C1826+660}

A further radio image of this object (Fig.\ 1) was taken to search for any
association of the three radio sources discussed in Paper I. None was
found, and henceforth they are treated as separate objects. A very faint 
identification has been found for 8C1826+660A (Fig.\ 3).

\section{analysis}

\subsection{The $R-z$ relation}

In Fig.\ 5 we plot the $R-z$ relation for the 8C galaxies and
quasars. 
Although such diagrams as this and the more famous $K-z$ relation
are somewhat ``blunt instruments'' for
studying radio galaxy evolution, where a number of different effects may be
contributing to the light as a function of redshift, they have some useful
features which can be exploited if they are used in conjunction with 
detailed studies of smaller samples of objects. 

As expected, the scatter in the $R-z$ relation is fairly low 
(with a few notable
exceptions) until $z\approx 0.6$, where the 4000\AA$\;$break moves
through the $R$-band. Above this redshift there is a wide range in
the amount of rest-frame UV light, as has been noted for the 3C 
sample (Lilly \& Longair 1984), but 
generally the galaxies are less luminous than
their 3C counterparts, consistent with the findings of 
Dunlop \& Peacock (1993) for their PSR sample. 

Up to at least $z=1$ there seems to be an upper envelope to the 
magnitudes which traces the
expected locus of an old stellar population, consistent with the idea 
that in many objects the host galaxy is not strongly evolving out to 
these redshifts. The presence of a weak alignment effect in the 7C
sample (Lacy et al.\ 1998a) suggests some of the light may be
influenced by the presence of the radio galaxy, however, so deep
spectroscopy of these objects is really needed to confirm this.

Six objects have no redshift information, and are plotted as open squares 
at the top left of
Fig.\ 5 with arbitary redshifts. Their most likely redshifts are in the 
range $z=1.2-1.8$ where no strong emission lines fall in the optical.
They mostly have faint optical magnitudes and so do not affect the discussion
below.

There is one interesting low-$z$ outlier on this plot. 8C1743+637
($R=21.3$; $z=0.324$) is in the cluster A2280, close to 8C1742+637, 
but was originally thought 
to be behind it from the point of view of both the radio structure (which 
is more FRII-like than FRI-like) and its magnitude (Paper II). It is 
possible that emission from a cluster galaxy is contaminating our 
spectrum, however. 

\begin{figure*}
\begin{picture}(400,400)
\put(-40,-130){\includegraphics{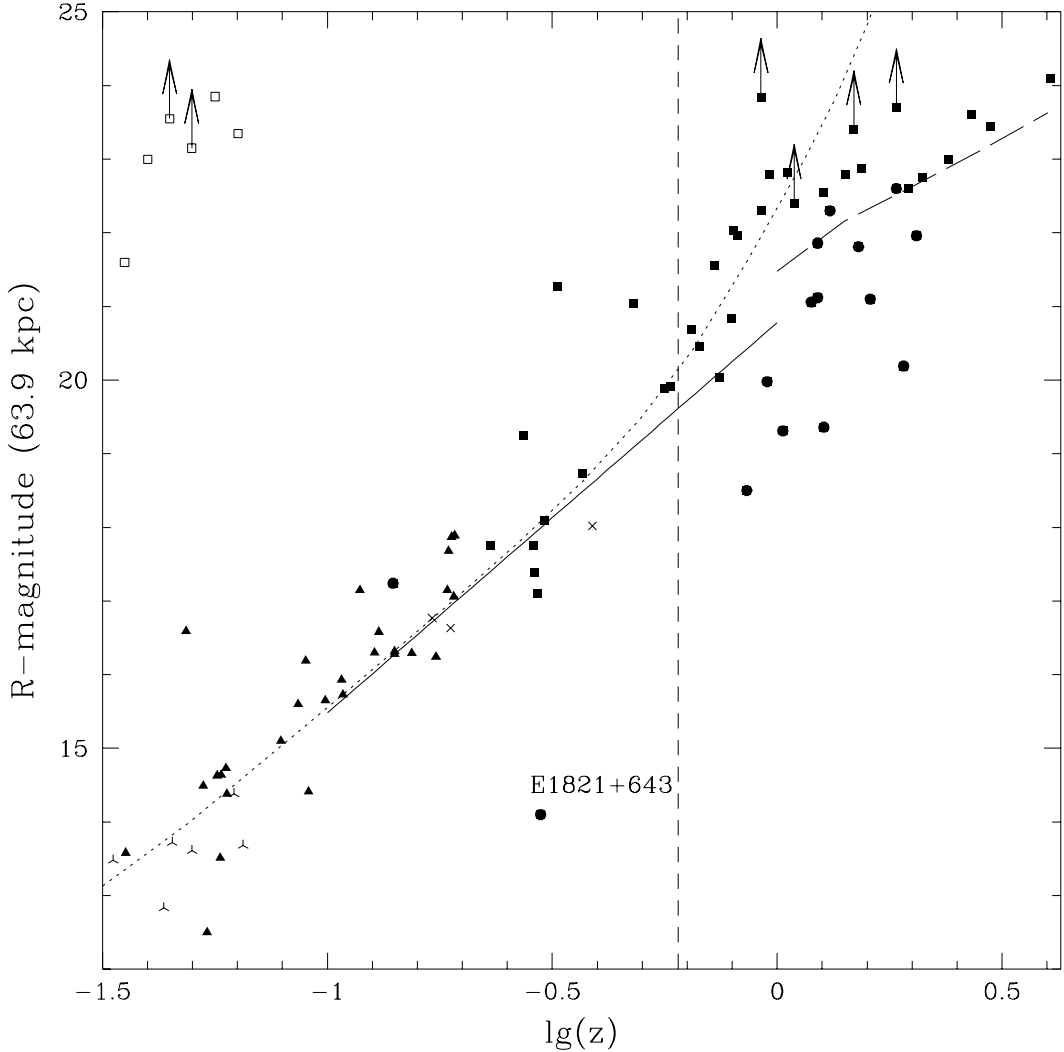}}
\end{picture}
\caption{The $R-z$ relation for the radio sources in the 8C sample, and 
the $z>0.03$ objects from the mostly 3C low-$z$ sample of Owen \& 
Laing (1989). The magnitudes have been corrected to a standard metric
aperture of 63.9 kpc according to the prescription of Eales et al.\ (1997). 
In the few cases where only an $I$-band magnitude has been measured (Table 5)
an $R-I$ of $0.5$, appropriate for a flat (in $f_{\lambda}$) spectrum has 
been assumed. The 8C objects are plotted as follows: solid squares represent 
narrow-line FRII radio galaxies, 
solid circles broad-line radio galaxies or quasars, and crosses represent FRI
radio galaxies. The Owen \& Laing sample has been plotted as triangles 
(FRII sources) or 3-pointed crosses (FRI sources). The cluster of open 
squares in the top left of the diagram represent 8C sources for which no 
redshift has been obtained,
in these cases an aperture correction appropriate to $z=1$ has been made.
The vertical dashed line is at $z=0.6$ where the 
4000\AA$\;$break is in the middle of the $R$-band. The dotted line 
is the expected $R-z$
relation for a non-evolving galaxy formed in a single burst at $z=20$
(using k-corrections from the Guiderdoni \& Rocca-Volmerange galaxy 
models). The long-dashed line is the $R$-band magnitude expected from 
a star formation rate of 65$M_{\odot}{\rm yr^{-1}}$ (see text). 
The solid line is the $R-z$ relation for 3C radio galaxies from 
Eales (1985).}
\end{figure*}

\subsection{$R-H$ colours}

Fig.\ 6 shows the $R-H$ colours for the sample. The low redshift
objects cluster around the no evolution prediction fairly
well. There are two very
red objects, 8C1743+645 with $R-H=4.7$, but an unknown redshift, and 
8C1816+671 with $R-H>5$ which has a (tentative) redshift of 0.92 based
on a single weak emission line (Paper IV). Presumably these are both 
$z\stackrel{>}{_{\sim}} 1$ objects with little blue light. 

\begin{figure}
\begin{picture}(200,200)
\put(-5,-65){\includegraphics{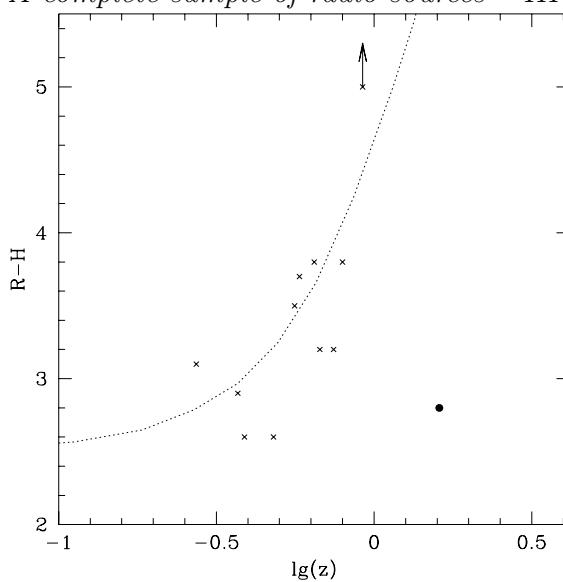}}
\end{picture}
\caption{$R-H$ versus redshift for a subset of the 8C sample. The objects
plotted are those radio source identifications with $R-H$-band colours 
in Table 4 and redshifts in Paper IV. Crosses represent
narrow-line radio galaxies and filled circles broad-line radio galaxies. The 
dotted line represents the non-evolving 1Gyr starburst model of Fig.\ 5.}
\end{figure}

\subsection{Discussion}

A remarkable property of the radio galaxies out to at least $z\approx 0.6$ 
is their ``standard candle'' nature over a wide range in radio 
luminosities, illustrated here in Fig.\ 5 and also present in 
other low radio luminosity, moderate redshift samples (e.g.\ Rixon, Wall 
\& Benn 1991). This has interesting implications for our understanding of 
what determines radio luminosity. To relate radio luminosity to the mass 
of the black hole, one could assume that radio jet power scales with
accretion power and therefore with black hole mass in the Eddington limit; 
thus one might expect less radio-luminous galaxies to have less massive black 
holes. Correlations between black hole mass and the mass of
spheroidal components in nearby galaxies (Kormendy \& Richstone 1995,
Magorrian et al.\ 1998), however, suggest that the small dispersion in 
host galaxy luminosities tranlates to a small dispersion in black hole mass.
Two possible explanations which would allow similar black hole masses
to produce a wide range of radio luminosities are: (1) the radio jet power is
constant for black holes of a given mass, and what determines the radio 
luminosity is the environmental density into which the jets propagate 
(Barthel \& Arnaud 1996); or (2) the radio jet power is determined by a factor 
which operates within the ``central engine'' region of galaxies with only the 
most massive black holes and which has to be linked 
in some way to the accretion power to produce the radio-luminosity --
emission-line luminosity correlation over a wide range in accretion 
rates (Willott et al.\ 1999). The lack of a strong correlation between 
radio luminosity and environment
(Hill \& Lilly 1991) and the correlation of radio-loud quasar optical and 
radio luminosities found by Serjeant et al.\ (1998) and Willott et al.\ 
(1998) both argue that the dominant factor in determining the radio 
luminosity is jet power, although environmental effects may of course play
a secondary r\^{o}le. 

The excess blue light above that expected from a non-evolving
elliptical seen in many objects at $z>2$ could clearly have a number
of origins, including ones associated with nuclear activity such as 
scattered quasar light and nebular continuum emission, but star
formation is also a likely contributor. Indeed, if, as widely
believed, radio galaxy hosts are giant ellipticals or their
progenitors at all redshifts, then one can estimate the amount of UV
light expected from star formation if radio galaxies form all their
$\approx 3\times 10^{11} M_{\odot}$ of stars in a single burst 
starting at $z=\infty$ and finishing at $z=1$. This is plotted on Fig.\ 5
as the long-dashed line, assuming a Scalo (1986) initial mass function. 
The average star formation rate, 
$65 M_{\odot}$yr$^{-1}$ is high enough that {\em unobscured}
star formation at or above this rate should be easily detected in our $z\sim 2$
objects. The fact that they all lie close to or fainter than this line could be
attributed to several causes: (1) obscuration by dust may be
extinguishing the UV light, (2) the star-forming episodes are of short 
duration, 
(3) the star-forming episodes occur mostly at very high redshift, 
(4) the star formation occurs in small
sub-units before the radio galaxy host assembles, or (5) the cosmology is 
significantly different from Einstein -- de Sitter.

Hughes and collaborators are performing a SCUBA-based submm survey of high
redshift radio galaxies from the 3C and 6C (Eales 1985) surveys
which should help to address point (1) (e.g.\ Hughes \& Dunlop 1998; 
Hughes et al.\ 1998). 
Current limits on star formation rates are about an order of 
magnitude higher than $65 M_{\odot}$yr$^{-1}$. Nevertheless, 
several detections of $2 \stackrel{<}{_{\sim}} z  \stackrel{<}{_{\sim}}4.3$ 
galaxies have already been made at this level. In addition Best et al.\ 
(1998) have reported a probable detection of the $z=1.2$ radio galaxy 3C324.
These detections, however, may only be upper limits too, in the sense
that much of the FIR emission could be coming from dust heated by the
AGN rather than from a starburst. The large dust masses involved, though, 
suggest at the very least that a large amount of star formation must 
have taken place in these objects by the time they are observed. 

Possibility (2) can probably be
ruled out unless the starbursts occur systematically before the
formation of the radio source. Even so, such very bright starbursts should 
be easily detected in the IR/submm (if they are dusty) or in optical 
narrow-band searches if not. 

Support for (3) or (4), very high redshift star 
formation either coherently or in small subunits which subsequently merge,
comes from the studies of the old stellar populations in the radio galaxies
3C65 ($z=1.2$; Lacy et al.\ 1995b; Stockton et al.\ 
1995) and 53W091 ($z=1.55$, Dunlop et al.\ 1996).
The old stellar populations seen in some radio galaxies 
contrast with the studies of Zepf 
(1997) and Barger et al.\ (1999) who find relatively low numbers 
of very red galaxies in optical/infrared studies of the Hubble Deep Field 
(HDF) and its flanking fields. These show
that only $\sim 50$ per cent of field ellipticals are in place by 
$z=1$, and that the remainder probably formed at $z\stackrel{<}{_{\sim}}3$.
The relatively high UV luminosity density produced by $z\sim 4$  
Lyman-dropout galaxies found by Steidel et al.\ (1998), however, points to the 
possibility that at least some galaxies formed at higher redshifts.
[Dust-obscured starbursts may affect this 
picture if they are important contributors to the star formation process
in ellipticals, but currently this is uncertain (e.g.\ Trentham, Blain \& 
Goldader 1998).]  

A possible explanation for the apparent discrepancy
between the ages of some radio galaxies and the field ellipticals is that
giant elliptical galaxies formed earlier than normal $\sim L_*$ elliptical 
galaxies. This would be broadly consistent with an
extrapolation of a trend deduced by Cowie et al. (1996) for more massive
galaxies to have their major episodes of star formation at earlier epochs.
If the oldest galaxies are the super-$L_*$ population, they will be rare in 
the relatively small area field surveys made to date.

Explanation (5), an incorrect cosmology, remains possible. For example an 
$\Omega_{\rm M}=0.5, 
\Omega_{\Lambda}=0.5$ cosmology with the same value of $H_0$ only requires 
a minimum star formation rate of $\approx 50 M_{\odot} {\rm yr^{-1}}$ to form 
all the stars by $z=1$, and the emission
will also appear fainter at a given redshift due to the higher luminosity 
distance.
These combine to make the $R$-magnitude for a $z=2$ galaxy forming stars at 
the minimum rate 0.8 mag fainter, for example.

The lack of UV emission also argues somewhat 
against jet-induced star formation being 
an important contributor to the star formation process of the galaxy as a 
whole, at least at $z\stackrel{<}{_{\sim}}3$. If it were, one would
expect all radio galaxies to show large amounts of UV and/or submm emission. 
It remains possible though that an initial episode of star formation 
activity could be 
triggered by an early outburst of radio jet activity at high redshift, and 
that subsequent episodes of radio jet activity would produce more modest 
starbursts, consistent with the relatively low limits on the unobscured 
star formation rates we infer for the 8C radio galaxies.

\section*{acknowledgements}
We thank the staff at the McDonald Observatory and the WHT for their 
assistance. The WHT is operated on the island of La Palma by
the Royal Greenwich Observatory in the Spanish Observatorio del
Roque de los Muchachos of the Instituto de Astrofisica de
Canarias. The VLA is operated by Associated Universities Inc.\ under 
a cooperative agreement with the National Science Foundation.


\begin{thebibliography}{99}

\bibitem{} Barger A.J., Cowie L.L., Trentham N., Fulton E.,
Hu E.M., Songalia A., Hall D., 1999, AJ, 117, 102

\bibitem{} Barthel P.D., Arnaud K.A., 1996, MNRAS, 283, 45

\bibitem{} Best P.N., Longair M.S., R\"{o}ttgering H.J.A., 1997, 
MNRAS, 292, 758

\bibitem{} Best P.N., R\"{o}ttgering H.J.A., Bremer M.N., Cimatti A., 
Mack K.-H., Miley G.K., Pentericci L., Tilanus R.P.J., van der Werf P.P., 
1998, MNRAS, in press (astro-ph/9809284)

\bibitem{} Cardelli J.A., Clayton G.C., Mathis J.S., 1989, ApJ, 345, 245

\bibitem{} Connolly, A.J., Szalay, A.S., Dickinson, M., Subbarao, M.U., 
Brunner, R.J. 1997, ApJ, 486, 11.    

\bibitem{} Cowie L.L., Songaila A., Hu E.M., Cohen J.G., 1996, AJ, 112, 839

\bibitem{} Dey, A., van Breugel, W., Vacca, W., Antonucci, R. 1997, 
ApJ, 490, 698

\bibitem{} Dunlop, J.S., \& Peacock, J.A. 1993, MNRAS, 263, 936

\bibitem{} Dunlop, J.S., Peacock, J.A., Spinrad, H., Dey, A.,
Jiminez, R., Stern, D. \& Windhorst, R.  1996, Nat., 381, 581

\bibitem{} Elias J.H., Frogel J.A., Matthews K., Neugebauer G., 1982, AJ, 87, 
1029

\bibitem{} Eales S.A., 1985, MNRAS, 217, 179

\bibitem{} Eales S.A., Rawlings S., Law-Green D., Cotter G., Lacy
M., 1997, MNRAS, 291, 593


\bibitem{} Hammer, F., Le Fevre, O., \& Angonin, M.C. 1993, 362, 324.


\bibitem{} Hill G.J., Lilly S.J., 1991, ApJ, 367, 1

\bibitem{} Hughes D.H., Dunlop J.S., to appear in Carilli et al., eds, 
``Highly Redshifted Radio Lines'', NRAO, Greenbank 9-11 October 1997. PASP,
(astro-ph/9802260)

\bibitem{} Hughes D.H., Dunlop J.S., Archibald E.N., Rawlings S., Eales S.A., 
1998, to be published in the proceedings of "The Birth of Galaxies", Xth 
Rencontres de Blois (astro-ph/9810253)


\bibitem{} Icke V., 1999, in, R\"{o}ttgering H.J.A., Best P.N., Lehnert M.D., 
eds, ``The most distant radio galaxies''.
Royal Netherlands Academy of Arts and Sciences, Amsterdam, p.\ 217

\bibitem{} Kolman M., Halpern J.P., Shrader C.R., Filippenko A.V., 1991, 
ApJ, 373, 57

\bibitem{} Kolman M., Halpern J.P., Shrader C.R., Filippenko A.V., Fink H.H., 
Schaeidt S.G., 1993, ApJ, 402, 514

\bibitem{} Kormendy J., Richstone D., 1995, ARA\&A 33, 581

\bibitem{} Lacy M., Rawlings S., Warner P.J., 1992, MNRAS, 256, 404 (Paper I)


\bibitem{} Lacy M., Hill G.J., Kaiser M.E., Rawlings S., 1993, MNRAS, 263, 707
(Paper II)

\bibitem{} Lacy M., Rawlings S., Eales S., Dunlop J.S., 1995b, MNRAS, 273, 821

\bibitem{} Lacy M., Riley J.M., Waldram E.M., McMahon R.G., Warner P.J., 
1995a, MNRAS, 276, 614

\bibitem{} Lacy M., Ridgway S.E., Wold M., Lilje P.B., Rawlings S., 1998a, 
MNRAS, in press

\bibitem{} Lacy M., Rawlings S., Hill G.J., Bunker A.J., Stern D., Ridgway 
S.E., 1998b, in press (Paper IV)

\bibitem{} Lacy M., Blundell K.M., Hill G.J., Kaiser M.E., Rawlings S., 1999c,
in preparation 


\bibitem{} Lilly, S.J. \& Longair, M.S. 1984, MNRAS, 211, 833.


\bibitem{} Madau, P. et al. 1996, MNRAS, 283, 1388

\bibitem{} Magorrian J., Tremaine S., Richstone D., Bender R., Bower G., 
Dressler A., Faber S.M., Gebhardt K., Green R., Grillmair C., Kormendy J., 
Lauer T., 1998, AJ, 115, 2285

\bibitem{} Owen F.N., Laing R.A., 1989, MNRAS, 238, 357


\bibitem{} Rees N., 1990, MNRAS, 244, 233


\bibitem{} Ridgway S.E., Stockton A.N., 1997, AJ, 114, 511

\bibitem{} Rixon G.T., Wall J.V., Benn C.R., 1991, MNRAS, 251, 243

\bibitem{} Scalo J.M., 1986, Fund.\ Cosmic Phys., 11, 1

\bibitem{} Serjeant,S., Rawlings,S., Maddox, S.J., Baker, J.C.,
Clements, D., Lacy, M. \& Lilje, P.B. 1998, MNRAS, 294, 494  

\bibitem{} Steidel C.C., Adelberger K.L., Giavalisco M., Dickinson M., 
Pettini M., 1998, ApJ, in press (astro-ph/9811399)

\bibitem{} Stockton A.N., Kellogg M., Ridgway S.E., 1995, ApJ, 443, L69

\bibitem{} Trentham N., Blain A.W., Goldader J., 1998, MNRAS, in press


\bibitem{} Willott C.J., Rawlings S., Blundell K.M., Lacy M., 1998, 
MNRAS, 300, 625

\bibitem{} Willott C.J., Rawlings S., Blundell K.M., Lacy M., 1999, 
MNRAS, in press 

\bibitem{} Zepf S.E., 1997, Nat, 390, 377

\end{thebibliography}
 \end{document}